# Out-of-Plane Resistance Switching of 2D $Bi_2O_2Se$ at Nanoscale

Wenjun Chen,[1,2] Rongjie Zhang,[1] Rongxu Zheng,[1] Bilu Liu[1,*]

[1]*Shenzhen Geim Graphene Center, Tsinghua-Berkeley Shenzhen Institute and Institute of Materials Research, Tsinghua Shenzhen International Graduate School, Tsinghua University, Shenzhen, 518055, P. R. China*

[2]*School of Electronic and Information Engineering, Foshan University, Foshan, 528000, P. R. China*

*To whom correspondence should be addressed. E-mail: bilu.liu@sz.tsinghua.edu.cn (BL)

**ABSTRACT:** 2D bismuth oxyselenide ($Bi_2O_2Se$) with high electron mobility shows great potential for nanoelectronics. Although in-plane properties of $Bi_2O_2Se$ have been widely studied, its out-of-plane electrical transport behavior remains elusive, despite its importance in fabricating devices with new functionality and high integration density. Here, we study the out-of-plane electrical properties of 2D $Bi_2O_2Se$ at nanoscale by conductive atomic force microscope. We find that hillocks with tunable heights and sizes are formed on $Bi_2O_2Se$ after applying vertical electrical field. Intriguingly, such hillocks are conductive in vertical direction, resulting in a previously unknown out-of-plane resistance switching in thick $Bi_2O_2Se$ flakes while ohmic conductive characteristic in thin ones. Furthermore, we observe the transformation from bipolar to stable unipolar conduction in thick $Bi_2O_2Se$ flake possessing such hillocks, suggesting its potential to function as a selector in vertical devices. Our work reveals unique out-of-plane transport behavior of 2D $Bi_2O_2Se$, providing the basis for fabricating vertical devices based on this emerging 2D material.

**KEYWORDS:** 2D materials; $Bi_2O_2Se$; nanoscale electrical property; out-of-plane resistance switching; unipolar conduction window

# 1. Introduction

Two-dimensional (2D) semiconducting bismuth oxyselenide ($Bi_2O_2Se$) with sizeable bandgap, high electron mobility, and excellent air stability is promising in high-performance nanoelectronics.[1-6] For example, $Bi_2O_2Se$ has been used to fabricate three-terminal memristors with the combination of short-term and long-term plasticity to realize neuromorphic functions.[7] In another work, gas sensors based on 2D $Bi_2O_2Se$ show high selectivity and ultralow oxygen detection limit of 0.25 ppm at ambient temperature because of the easy trap of oxygen by the Se vacancies on the surface of $Bi_2O_2Se$.[8] Furthermore, controlled thermal oxidation of $Bi_2O_2Se$ is used to directly construct dielectric/semiconductor ($Bi_2SeO_5$/$Bi_2O_2Se$) structures like in the case of silicon ($SiO_2$/Si). Because the dielectric constant of the $Bi_2SeO_5$ layer is high as ~21, such $Bi_2SeO_5$/$Bi_2O_2Se$ structures are promising for fabricating high performance electronics.[2] Additionally, the $Bi_2O_2Se$-based photodetectors with decent performance in terms of on/off ratio ($\approx 10^9$),[9] photodetectivity ($3.4 \times 10^{15}$ Jones),[9] broadband detection (360-1800 nm),[10] and photoresponse time ($\approx 1$ ps)[3] have been achieved. The above recent achievements suggest that the devices based on $Bi_2O_2Se$ are important components in future integrated 2D electronics and optoelectronics. Note that all these applications are based on the charge transport in the horizontal direction of $Bi_2O_2Se$ at micro-level. It is therefore intriguing to exploit the out-of-plane transport behavior of 2D $Bi_2O_2Se$ to extend its functionality.

Besides microscale charge transport, the nanoscale electrical properties of 2D materials play a significant role in the performance and/or mechanism of related microscopic or macroscopic electronic devices.[11] Conductive atomic force microscope (CAFM) is a useful tool with high spatial resolution to study the electrical properties of 2D materials at nanoscale.[12] Along this direction, the performance of 2D dielectric layers has been recently revealed by CAFM technique. For example, Hattori et al. have shown that 2D BN is subjected to a layer-by-layer breakdown with a medium dielectric strength of ≈ 12 MV cm$^{-1}$,[13] while 2D CaF$_2$ possesses a high dielectric strength of ≈ 27.8 MV cm$^{-1}$ owing to its cubic lattice structure.[14] Researchers also used CAFM to reveal the vertical electronic properties of nanoscale domains and domain boundaries in strong coupled systems such as twisted bilayers of graphene[15] and transition metal dichalcogenides.[16-17] Moreover, CAFM was used to monitor the formation of conductive filaments in memristors based on 2D materials to reveal the operating mechanism. For example, the resistance switching generated in 2D TiO$_2$ relies on the electromigration of the internal oxygen vacancies in the vertical direction,[18-19] while the memory behaviors of 2D BN stems from the defects generated during the synthesis process.[20-21] Although some intrinsic physical properties of 2D Bi$_2$O$_2$Se, such as mechanical properties[22] and ferroelectricity and piezoelectricity,[23] have been uncovered by scanning probe techniques very recently, its out-of-plane electrical properties remains elusive, despite its significance in developing devices with new functionality and meeting the requirements of high integration density and compatibility.[24-25]

Herein, we report unique out-of-plane electrical properties of 2D Bi$_2$O$_2$Se. We find that hillocks are formed on 2D Bi$_2$O$_2$Se after applying vertical electrical fields. The heights and widths of the hillocks and accordingly the electrical performance of the 2D Bi$_2$O$_2$Se can be controlled by the amplitude, cycle numbers of the electrical field, and thickness of materials. Moreover, nanoscale conductive pathways form at the hillock locations, making the initially insulting Bi$_2$O$_2$Se exhibit out-of-plane resistive switching function and ohmic feature in thick and thin flakes, respectively. Such resistance switching devices show transformation from bipolar to stable unipolar conduction window, enabling its use as a selector in vertical devices. This study not only uncovers the out-of-plane electrical transport behavior of 2D $Bi_2O_2Se$ for the first time, but also opens a door to explore its use in nanoelectronics with new functions.

## 2. Results and Discussion

2D $Bi_2O_2Se$ were grown by vapor deposition method[9] and were transferred onto highly-conductive substrate made of Au/Cr coated $SiO_2$/Si wafer (hereinafter referred to as Au substrate for short)[22] (See growth and transfer details in *Supporting Information*). The as-synthesized flakes are rectangular single crystals with different thicknesses (Figure 1a) with high quality, as indicated by the $A_{1g}$ mode (~ 159 $cm^{-1}$) of Raman spectrum (blue curve in Figure S1). The samples have uniform structure and composition, as shown by optical microscope (Figure 1b) and the intensity distribution of the $A_{1g}$ peak in corresponding Raman map (Figure 1c). After transfer onto the target substrate, $Bi_2O_2Se$ flakes still keep the initially rectangular shape (Figure 1d). The high-magnification optical microscope image (Figure 1e) and corresponding Raman map of the $A_{1g}$ peak intensity (Figure 1f) of the transferred $Bi_2O_2Se$ flakes jointly show its good uniformity and similar quality with the as-grown ones. In addition, XPS spectra indicate the chemical composition of $Bi_2O_2Se$ flakes (Figure 1g-i), which is consistent to the results reported previously.[9, 22] These characterization results above together show the high quality and structural integrity of $Bi_2O_2Se$ flakes after transferring on conductive Au substrate, providing basis for subsequent electrical studies.

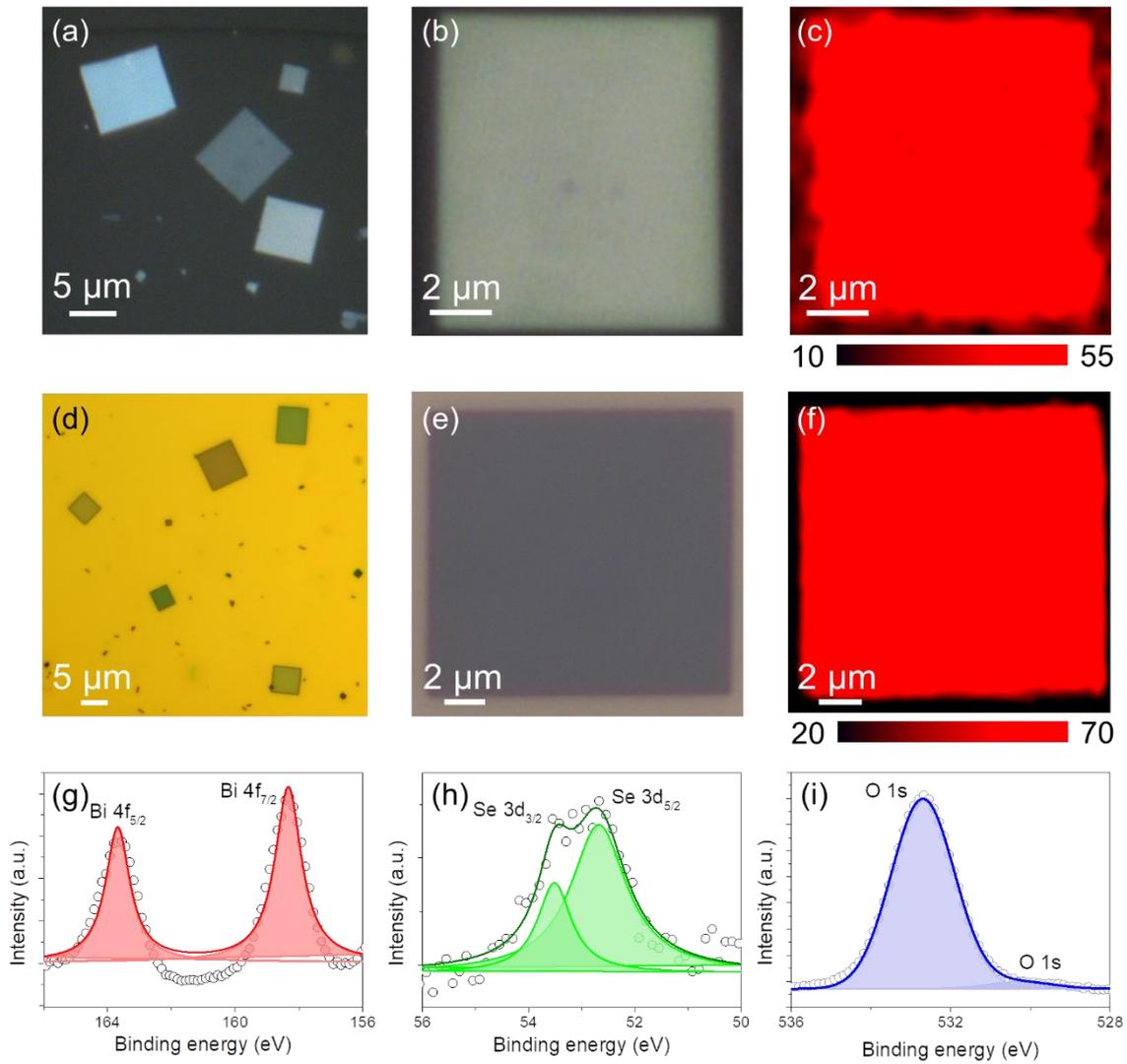

**Figure 1.** Transfer of 2D $Bi_2O_2Se$ onto conductive Au substrate and characterization. (a) Optical microscope image of as-grown $Bi_2O_2Se$ on mica. (b) High-magnification optical microscope image and (c) corresponding Raman $A_{1g}$ intensity map of $Bi_2O_2Se$ on mica. (d) Optical microscope image of $Bi_2O_2Se$ after transferring on conductive Au substrate. (e) High-magnification optical microscope image and (f) corresponding Raman $A_{1g}$ intensity map of $Bi_2O_2Se$ on Au substrate. XPS spectra of $Bi_2O_2Se$ on Au substrate with (g) Bi 4f, (h) Se 3d and (i) O 1s peaks.

After transferring 2D $Bi_2O_2Se$ onto conductive Au substrate (Figure 2a), CAFM is used to study its out-of-plane transport behavior. Figure 2b displays the CAFM setup where the Au substrate is one of the electrodes, and AFM cantilever serves as the other one to apply electric field on $Bi_2O_2Se$ samples. Prior to CAFM measurements, AFM scanning was performed to image the microscopic morphology and measure the thickness of the selected $Bi_2O_2Se$ flake (Figure S2). Afterwards, three random positions on the flake, as marked by dotted circles with different colors in Figure 2c, were chosen and cyclic triangular wave electric field with different amplitudes was applied in the vertical direction by the CAFM tip. Interestingly, we found that a hillock with a lateral size of ~300 nm formed at the local place where the tip and $Bi_2O_2Se$ flake contacted under the cyclic voltage with an amplitude of 10V (red dotted in Figure 2c and 2d). Similar phenomenon, i.e., the formation of a hillock after applying electrical field, was observed at the position that was stimulated by the electric field with the same cycle number but a lower amplitude of 5V (green dotted in Figure 2c and 2e). The height of this hillock is lower than that formed under the 10 V sweeping voltage. However, for the position simulated by the electric field with the amplitude of 2V, the position keeps flat without the formation of a rough structure (blue dotted in Figure 2c and 2f). It is further analyzed that the heights of the hillocks formed under 30 cycles of the sweeping voltage with the amplitude of 10 and 5V are 18.8 and 4.5 nm respectively (Figure 2g), which demonstrate a positive correlation between the heights of hillocks and the amplitude of the electric fields. In order to trace the formation process of the hillocks, the height at the same position was measured after each cyclic application of electric field with the amplitude of 10V. We found that the hillock forms (~2.5 nm) since the first sweep of the applied electric field, and the height increases with increasing cycle number, as exhibited in Figure 2h. Then the height of the hillock keeps stable as ~19 nm after 5 sweeps of the electric field (Figure 2i). Cross-sectional high-angle annular dark-field (HAADF) imaging and Energy Dispersive Spectrometer (EDS) analysis (Figure S3) were performed to clarify that the content of oxygen on the surface of the hillock is higher than that at the internal areas. Accordingly, it is reasonable to deduce that the region where electric field was applied was oxidized in air, causing the increase of local layer spacing[2] and formation of hillock on $Bi_2O_2Se$.

In contrast, the mechanism of similar phenomena in other 2D materials such as multilayer h-BN was proposed as its breakdown under strong electrical field.[20] Taken together, the above results show that protruded structure can be constructed at pre-defined positions on $Bi_2O_2Se$ flakes *via* the precise location of the CAFM tip. In addition, the height of the hillock is facilely tuned by modifying the amplitude and cycle number of the applied electric field.

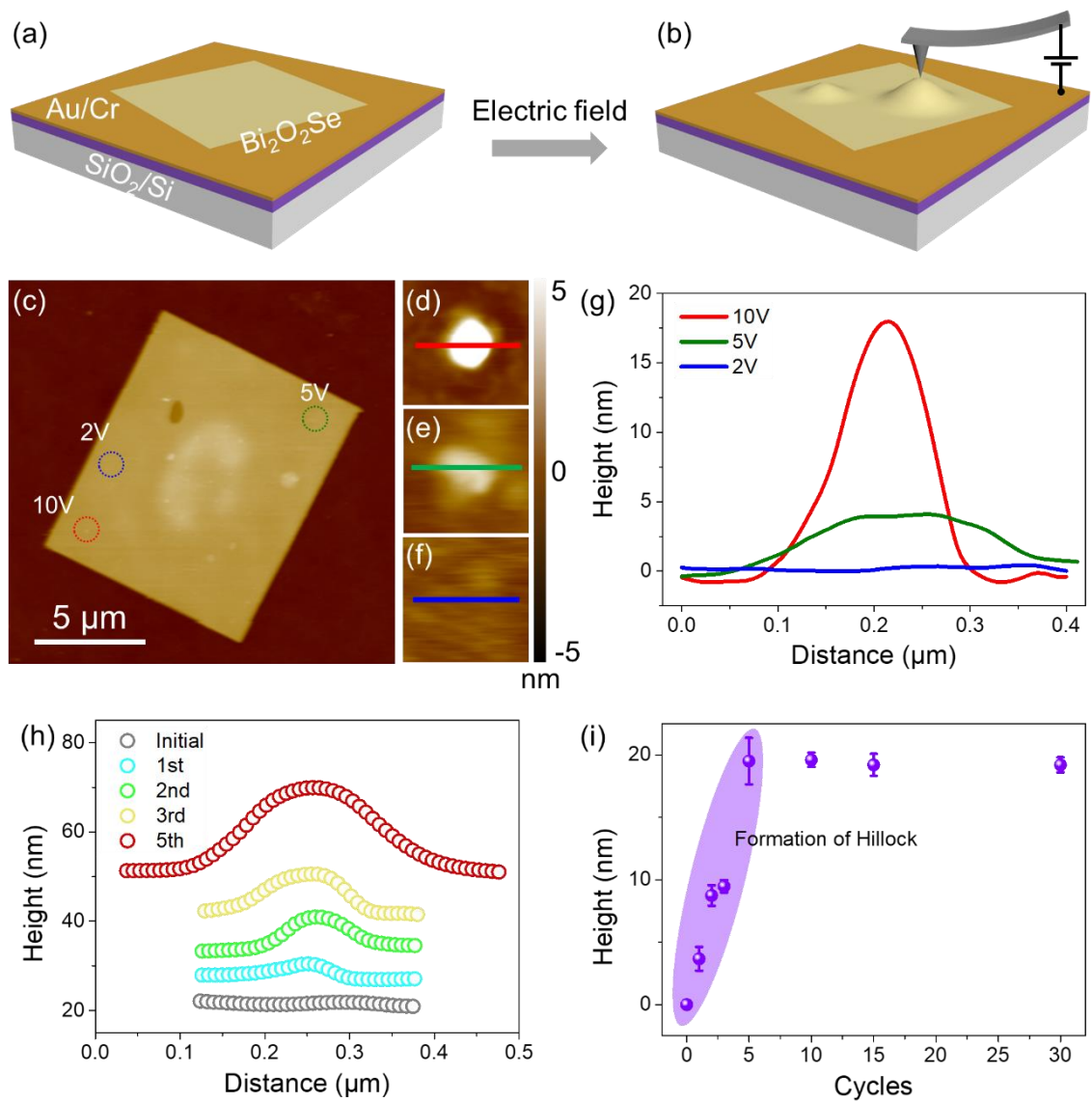

**Figure 2.** Formation of hillocks on 2D $Bi_2O_2Se$. Schematic of (a) $Bi_2O_2Se$ on Au substrate and (b) CAFM apparatus set to apply vertical electric field on $Bi_2O_2Se$. (c) AFM image of a $Bi_2O_2Se$ flake. Three positions marked by red,

green and blue dotted circles were set with 30-cycle sweep of triangular electrical field with the amplitude of 10, 5, and 2V, respectively. (d-f) Zoom-in AFM images with a color bar of the three positions in (c). (g) Height profiles corresponding to red line in (d), green line in (e), and blue line in (f), respectively. (h) Height of the hillock in different cycles and (i) change of the heights with increasing cycle number under the cyclic triangular electric field with the amplitude of 10V. Error bar in (i) was calculated by the height of three hillocks formed after the same cycles of the cyclic electric field.

Electric field in the vertical direction not only triggers the formation of hillocks on $Bi_2O_2Se$, but also generates out-of-plane conduction. As marked by the red dotted rectangular in Figure 3a, the "forming zone" was formed during the AFM scanning process that the probe along with the application of 5V constant electric field initiates the formation of hillocks. The rest part of the $Bi_2O_2Se$ flake does not exhibit morphology changes because the electric field is removed. Importantly, after the forming process, the hillock positions became more conductive that nanoscale conductive pathways in the vertical direction were identified and marked by the yellow dotted rectangular in the corresponding CAFM map (Figure 3b). The current of the nanoscale conductive paths was about 1 nA under the bias of 2V (top in Figure 3c), while the rest area is very resistive with current of ~1 pA (bottom in Figure 3c). As shown in Figure 3d, our AFM studies show that 70-nm-thick $Bi_2O_2Se$ flake is at insulating state originally (blue plots). Then it becomes conductive (green plots) under the application of cyclic triangular wave electric field with the amplitude of 10V, which demonstrates the forming process of conductive channels. With the increasing cycle number to 30$^{th}$, the *I-V* plots with two conduction windows under both positive and negative electric field of this flake (red plots) clarifies out-of-plane resistance switching of the 70-nm-thick $Bi_2O_2Se$. After the forming process, the resistance switching can operate under a lower electric field (Figure S4).

Apart from the cycle number, the amplitude of the sweeping electric field also contributes to the out-of-plane resistance switching phenomenon of $Bi_2O_2Se$. Specifically, resistance switching forms under 30-cycle sweep of the electric field with the amplitude of 10V (red plots in Figure 3e). Although the periodic voltage with the amplitude of 5V makes the $Bi_2O_2Se$ flake conductive at the $30^{th}$ cycle (blue plots in Figure 3e), conduction windows are difficult to form in this case. Additionally, no electrical responses in the vertical direction was collected under a 2V electric field (grey plots in Figure 3e). By combination with the morphological analyses in Figure 2, we conclude that the formation of conduction windows is related to the height of hillocks. The thickness of $Bi_2O_2Se$ flake is another factor that imposes effects on its out-of-plane electrical features. The transformation from insulating to conductive state combined with the formation of hillocks (Figure S5) were also monitored in a thinner $Bi_2O_2Se$ flake with a thickness of 8 nm under 10 V sweeping electric field. Different from the thick sample, the *I-V* plots of this 8-nm-thick flake show ohmic characteristics instead of resistance switching after the application of 20 cycles of the voltage suppress, which stems from the electrically hard breakdown of the thin sample (Figure 3f). The *I-V* plots during the same cycle of various $Bi_2O_2Se$ flakes with different thicknesses were further analyzed and the results show that conductive channels are easier to form in thinner sample under the same electrical field amplitude (Figure 3g). Therefore, the out-of-plane nanoscale electrical properties of 2D $Bi_2O_2Se$ show a unique thickness-dependent behavior, i.e., thick flakes show resistance switching phenomenon while thin ones show linear *I-V* curves.

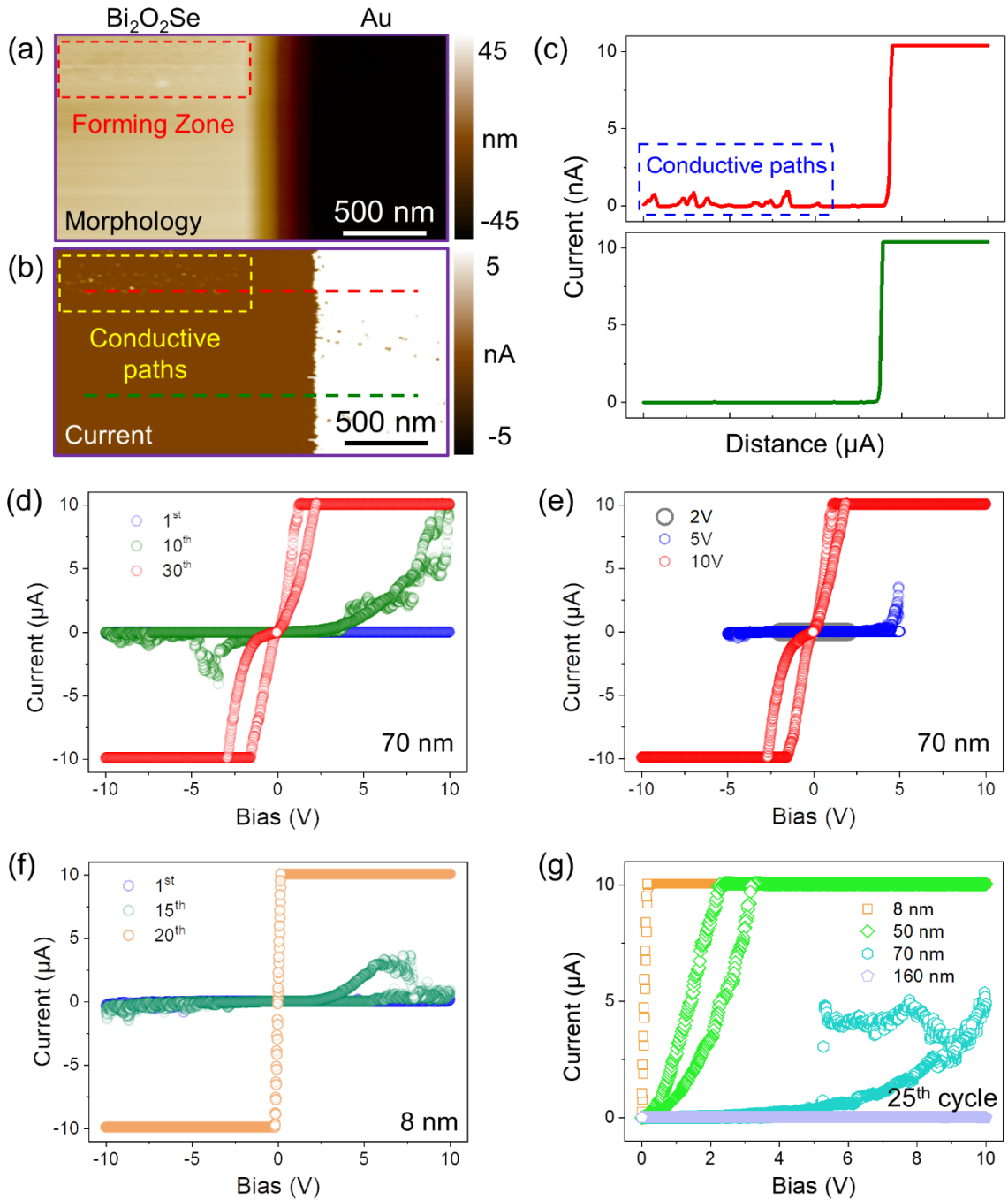

**Figure 3.** Out-of-plane resistance switching of 2D $Bi_2O_2Se$. (a) AFM Morphology and (b) corresponding CAFM current map at the edge of a $Bi_2O_2Se$ flake. The forming zone marked by red dotted rectangular in (a) was imaged with the application of a 5V bias. The conductive pathways marked by yellow dotted rectangular in (b) were formed under the application of a 2V bias. (c) Current analyses of the red and green profiles in (b). The formation of conductive paths with a current of ~ 1 nA is marked by the blue dotted rectangular. (d) *I-V* plots in different cycles of $Bi_2O_2Se$ with a thickness of 70 nm under the application of 10V cyclic triangular electrical field. (e) *I-V* plots in the 30[th] cycle of $Bi_2O_2Se$ with a thickness of 70 nm under the application of cyclic triangular electrical field with different amplitudes. (f) *I-V* plots in different cycles of $Bi_2O_2Se$ with the thickness of 8 nm under the application of 10V cyclic triangular electrical field. (g) *I-V* plots in the 25$^{th}$ cycle of $Bi_2O_2Se$ with different thicknesses under the application of cyclic triangular electrical field with the amplitude of 10V.

Next, we study the polarity of the resistance switching behavior of $Bi_2O_2Se$. Essentially, the mode of out-of-plane resistance switching in $Bi_2O_2Se$ flakes experiences a change during the sweep of cyclic, which is indicated by three *I-V* curves collected sequentially with CAFM. To start with, the *I-V* curve shown in Figure 4a claims the bipolar conductive behavior of $Bi_2O_2Se$ after the formation of conductive paths in the vertical direction, which is consistent with most of the memory devices based on 2D materials.[26] Under the 0 to 10V forward sweep (Step 1), the flake was at high-resistance (HR) state, and then switched to low-resistance (LR) state during the sweep from 10 to 0V (Step 2). Afterwards, the LR state was maintained under the negative electric field swept from 0 to -10V (Step 3) and restored to HR under the -10 to 0V sweep (Step 4). During the next cycle, a sudden transformation from HR to LR state was observed under the forward sweep of the positive voltage (Step 2 in Figure 4b). Subsequently, the electrical properties still comply with LR (0 to -10V, Step 4) to HR (-10 to 0V, Step 5) switching under the sweep of negative voltage. From then on, the *I-V* curves exhibit alternative switching between LR and HR state (Figure 4c), demonstrating a unipolar conductive behavior which is converted from initial bipolar resistance switching. The alternation between resistance states is also found in the in-plane[27] and out-of-plane[28] memory devices based on other 2D materials. The cyclic CAFM measurements in Figure 4d further point out the high stability of the unipolar resistance switching of $Bi_2O_2Se$ in the vertical direction. In comparison to other 2D oxyselenides with out-of-plane unipolar resistance switching behavior such as $HfSe_{2-x}O_x$, $Bi_2O_2Se$ possesses lower set/reset voltage.[29] This unipolar resistance switching of $Bi_2O_2Se$ may result from the dynamical balance between the formation of current paths and the production of Joule heat that Joule heat makes the internal vacancies of $Bi_2O_2Se$ move to produce current, which feeds back to generate Joule heat.[28] With the combination of EDS map shown in Figure S3 and the above discussions, the temperature of the conductive region is dramatically increased to trigger local oxidation and interlayered expansion that inspires the formation of hillock.

In order to further verify this hypothesis, two random positions of a flake were selected to apply the same cyclic in air and $N_2$ gas, respectively. In air, we found a hillock formed at the place to make resistance switching happen. As a sharp contrast, in $N_2$, no morphology change on $Bi_2O_2Se$ was observed and it kept electrically insulating (Figure S6). In consequence, the existence of $O_2$ is responsible for the morphological changes and resistance switching of $Bi_2O_2Se$ under electrical field. The out-of-plane resistance switching assures $Bi_2O_2Se$ for applications in future nanoelectronics with new functionality, high integration and high compatibility. In addition, crossbar electronic devices are subjected to the sneak current in the vertical direction, which can be mitigated by the addition of selector device with unilateral electrical conduction between two layers of electrodes.[30] Therefore, owing to the out-of-plane unipolar conduction behavior, $Bi_2O_2Se$ is suitable for applications in vertically-oriented devices as a functional material as well as a selector.

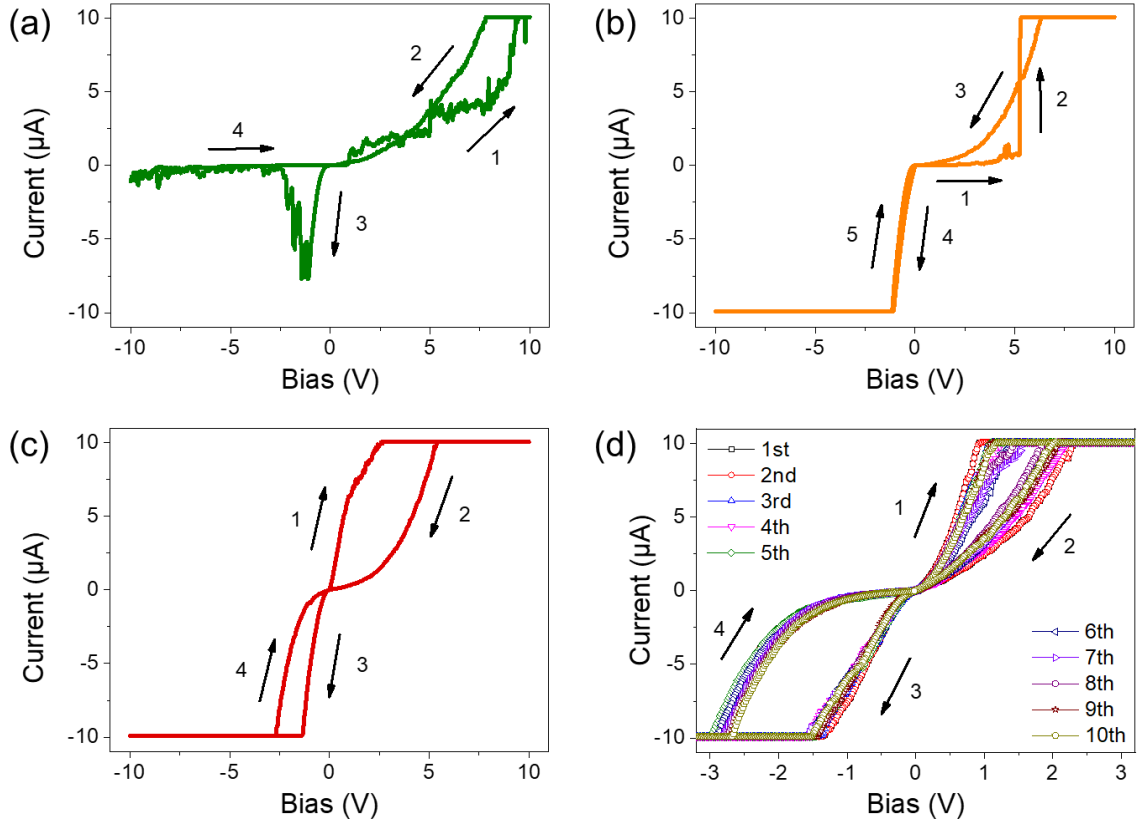

**Figure 4.** Transformation from bipolar to unipolar conduction window in $Bi_2O_2Se$ devices. (a-c) Three sequential *I-V* curves collected under the application of 10 V cyclic triangular electrical field. (d) 10-cycle measurements of the resistance switching of the same $Bi_2O_2Se$ flake.

## 3. Conclusion

We have revealed the out-of-plane electrical properties of 2D $Bi_2O_2Se$ at nanoscale for the first time. We found that hillocks formed on 2D $Bi_2O_2Se$ under applying cyclic triangular electrical field. The height of these hillocks are controlled by amplitude or cycle number of the cyclic voltage suppress and the thickness of $Bi_2O_2Se$ flakes. Noteworthy, the hillock positions contain nanoscale conductive pathways, leading to a unique vertical resistance switching behavior in thick $Bi_2O_2Se$ flakes while ohmic characteristic in thin ones. In addition, the devices based on thick $Bi_2O_2Se$ flakes exhibited transformation from bipolar to stable unipolar resistance switching behavior. The work not only discloses novel out-of-plane transport behaviors of 2D $Bi_2O_2Se$, but also opens an avenue for the use of 2D materials in highly-integrated nanoelectronics with new functionality.

## 4. Experimental Section

*Growth, Transfer, and Characterization of 2D $Bi_2O_2Se$*: 2D $Bi_2O_2Se$ samples were synthesized on mica substrate by vapor-solid deposition,[9] and were transferred onto Au/Cr-coated $SiO_2$/Si substrate by the PDMS-mediated[22] or PMMA-assisted method.[9] For preparation of Au/Cr-coated substrate, $SiO_2$/Si substrate was put into the chamber of electron beam evaporation system (TSV-1500, Tianxingda Vacuum Coating Equipment Co., Ltd., China). Then, Cr layer with a thickness of 5 nm was deposited on the $SiO_2$/Si substrate at the speed of 0.2 Å $s^{-1}$, which was followed by the deposition of Au layer with the thickness of 40 nm at the speed of 2 Å $s^{-1}$, in order to obtain conductive Au/Cr/$SiO_2$/Si substrate.

The surface morphology of $Bi_2O_2Se$ samples was characterized by optical microscope (Imager A2m, Carl Zeiss, Germany). XPS (ESCALAB 250Xi, Thermo Fisher, USA) was used to reveal the chemical composition of $Bi_2O_2Se$ flakes. Raman spectrometer (HR800, Horiba JY, Japan) was used to collect the Raman spectra and mapping of $Bi_2O_2Se$. The spot size of the incident laser with the wavelength of 633 nm is 500 nm, and each step of Raman mappings is 1 μm. The cross section of hillocks on $Bi_2O_2Se$ was obtained by focused ion beam (Helios UC, FEI, USA). HAADF imaging and EDS analysis were performed to characterize the chemical composition of hillocks on $Bi_2O_2Se$ with a TEM (Talos, FEI, USA) at an operating voltage of 200 kV.

*CAFM Measurement of out-of-plane conductance of Bi$_2$O$_2$Se*: AFM scanning (Cypher ES, Oxford Instruments, USA) was performed to image the morphology and measure the thickness of Bi$_2$O$_2$Se with contact mode. The AFM probe (FM-LC, Adama Ltd., Ireland) with doped diamond coating and the radius of 20 nm was used for morphology imaging, application of cyclic triangular electric field in the vertical direction at the frequency of 1 Hz, and collection of *I-V* curves in CAFM setup. The compliant current was 10 µA. For obtaining CAFM map of the out-of-plane current, a bias was set and applied by the tip while scanning the morphology of samples. For the measurements in pure N$_2$, which is introduced into the cell for 10 min to eliminate the air, eventually the pressure of N$_2$ maintained at 190 mbar. The frequency of the sweeping voltage is 1 Hz.

**Supporting Information**

Supporting Information is available from the Wiley Online Library or from the author.

**Acknowledgements**

The authors acknowledge the supports by the National Natural Science Foundation of China (Nos. 51920105002, 51991340, and 51991343), the Bureau of Industry and Information Technology of Shenzhen for the "2017 Graphene Manufacturing Innovation Center Project" (No. 201901171523), the Shenzhen Basic Research Project (Nos. JCYJ20200109144620815 and JCYJ20200109144616617), and the Guangdong Innovative and Entrepreneurial Research Team Program (No. 2017ZT07C341).

**Conflict of Interest**

The authors declare no conflict of interest.